# UBe$_{13}$ and U$_{1-x}$Th$_x$Be$_{13}$ – Unconventional Superconductors

G. R. Stewart

Department of Physics, University of Florida, Gainesville, FL 32611

**Abstract:** UBe$_{13}$ was the second discovered heavy fermion superconductor, and numerous pieces of evidence exist that imply that it is an unconventional (non-BCS s-wave) superconductor. Exhibiting even more signs of unconventional superconductivity, Th-doped UBe$_{13}$ is perhaps the most puzzling of any of the unconventional superconductors. This review considers both the parent, undoped compound as well as the more interesting U$_{1-x}$Th$_x$Be$_{13}$. After summarizing the rather thorough characterization – which because of the interest in these compounds, has continued from their discovery in 1983 and 1984 to date - these properties are compared with a recent 'template' for determining whether a superconductor is unconventional. Finally, further experiments are suggested.

## I. Introduction

Bucher et al. [1], in a study of 16 different arc-melted polycrystalline $MBe_{13}$ compounds, reported in 1975 that $UBe_{13}$ exhibited superconductivity at a 'sharp' transition at 0.97 K in the ac susceptibility. Contrary to their expectation that the superconductivity was filamentary and would be destroyed by grinding, Bucher et al. found that grinding did not diminish or shift the superconducting transition signal. By applying 6 T, they found that $T_c$ decreased by 0.3 K, giving a $\Delta H_{c2}/\Delta T$ slope near $T_c$ of -20 T/K. Unfortunately for an early start to the study of heavy fermion superconductivity (HFS), discovered in $CeCu_2Si_2$ four years later, Bucher et al. only measured the specific heat (which would have shown bulk superconductivity at $T_c$) down to 1.8 K as shown in Fig. 1. (Note the excellent agreement between Bucher et al.'s early specific heat data and later measurements.) The upturn in C/T below 5 K seen in Bucher et al.'s data was in fact the first indication ever seen of heavy fermion (high effective mass m*) behavior, although not recognized as such at the time. Such large C/T values at low temperatures, the defining measurement for "heavy fermion" behavior, were next seen (and first correctly recognized as caused by an extremely high electronic density of states due to strong correlations) in non-superconducting $CeAl_3$ almost exactly one year later, at the end of 1975. [2] (For a discussion of the theory of how 4f and 5f electrons and their correlations can create such large m*, see, e. g., refs. 3-4.)

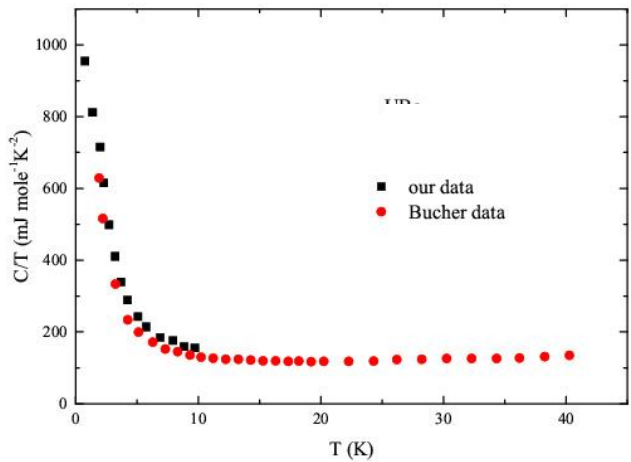

Fig. 1 (color online): Normal state specific heat divided by temperature (C/T) data of $UBe_{13}$ down to 1.8 K from Bucher et al. [1] (red circles) and down to 1.05 K from Kim et al. [5] (black squares) show the rapid upturn in C/T at low temperatures that is characteristic [6] of many heavy fermion materials. The strong temperature dependence of C/T is <u>prima facie</u> evidence of non-Fermi liquid behavior, which has been confirmed [7] by, e. g., non-Fermi liquid temperature dependences in both C/T and the resistivity.

In 1983, Ott et al. [8] – benefitting from a thorough exploration of HFS in $CeCu_2Si_2$ by the group of Steglich [9] and others – reported bulk superconductivity in $UBe_{13}$. Using single crystal $UBe_{13}$ grown from Al flux, Ott et al. reported C/T (T→0) as 1100 mJ/molK$^2$ (compare the data down to 1.05 K in Fig. 1) and $dH_{c2}/dT|_{Tc}$ = -25.7 T/K, with a resistively determined $T_c$ of 0.86 K. More importantly, Ott et al.'s data show a bulk anomaly ΔC in the specific heat starting at 0.9 K and peaked at 0.7 K, proving that the superconductivity in $UBe_{13}$ is a bulk phenomenon.

Approximately one year after Ott et al.'s discovery of HFS in $UBe_{13}$, while doping eight different elements (M) onto the U site in $UBe_{13}$ ($U_{1-x}M_xBe_{13}$) and tracking $T_c$ suppression with composition, Smith et al. [10] discovered a non-monotonic depression of $T_c$ with increasing x in polycrystalline samples of $U_{1-x}Th_xBe_{13}$, see Fig. 2. Such a non-monotonic variation of $T_c$ with doping concentration is highly unusual.

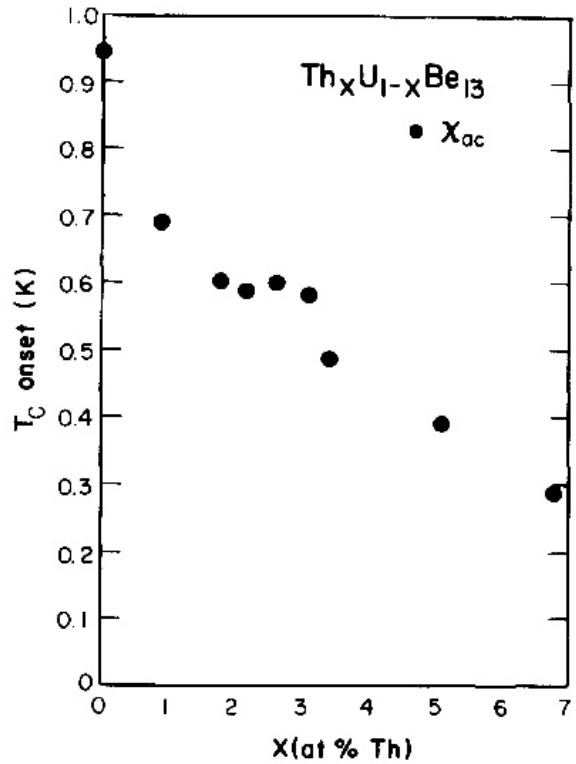

Fig. 2: ac magnetic susceptibility [10] of Th-doped $UBe_{13}$, showing that between 1.8 and 3% Th substitution the $T_c$ remains approximately constant. Smith et al. [10] also studied seven other M dopants in $U_{1-x}M_xBe_{13}$ for their effect of $T_c$. For approximately a concentration of x=0.018, La, Ce, and Np showed about the same, 0.35 K, suppression of $T_c$ as Th, while Ba and Sc showed less suppression and Gd more. 1.6 % Lu suppressed $T_c$ below 0.045 K. Only Th in $UBe_{13}$ is known to have the non-monotonic variation of $T_c$ pictured here.

Soon thereafter, Ott et al. [11] – in polycrystalline samples – showed that in the composition range $0.022 \leq x \leq 0.038$ in $U_{1-x}Th_xBe_{13}$ there exists either a distinct *second peak* in the specific heat ($0.026 \leq x \leq 0.033$) or at least a clear shoulder (x=0.022, 0.038). (It should be noted that the intercomparability of $T_c^{onset}$ for both transitions in $U_{0.970}Th_{0.030}Be_{13}$ (nominal concentration of arc-melted material) between the work of Ott et al. [11] and the later work of Kim et al. [12], appears to be quite good. The overall reproducibility between laboratories for concentrations given for x in $U_{1-x}Th_xBe_{13}$ to two significant figures will be discussed further below.) Specific heat data on high purity polycrystalline samples to delineate this behavior will be presented below. As an introduction, Fig. 3 presents the specific heat data in both the superconducting and normal states by Kim et al. [12] for both $UBe_{13}$ and $U_{0.97}Th_{0.03}Be_{13}$.

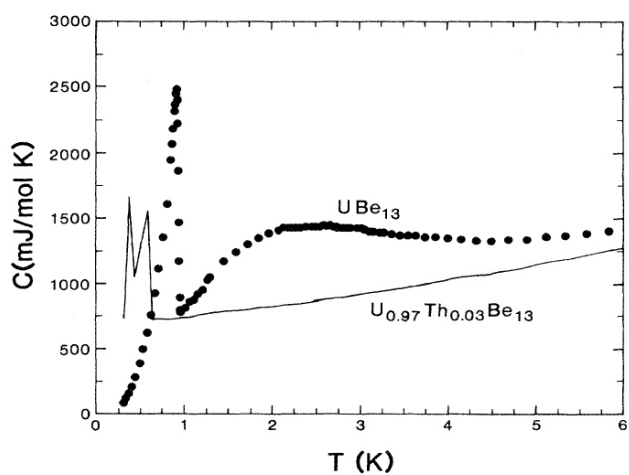

Fig. 3. Data [12] down to 0.3 K on two high purity, annealed polycrystalline samples: $UBe_{13}$ (annealed at 1200 C for 3.5 weeks) and $U_{0.97}Th_{0.03}Be_{13}$ (annealed at 1400 C for 7.3 weeks), with the latter showing a double peak structure in the bulk in the composition range which Smith et al. [10] found the onset $T_c$ approximately independent of Th concentration.

The cause of this second feature in the specific heat of Th-doped $UBe_{13}$ was discussed, and one of the possibilities mentioned [11] was that the second anomaly "indicates a continuous phase transition from one superconducting state below $T_{c1}$ to another below $T_{c2}$," i. e. unconventional superconductivity. Despite thorough investigation into other dopants (see, e. g., Smith et al. ref. 10), Th remains the only example of a dopant in $UBe_{13}$ causing a second transition.

It is the goal of this review to summarize the large amount of experimental work on $UBe_{13}$ and $U_{1-x}Th_xBe_{13}$ to date, and discuss the evidence in both for unconventional superconductivity. Several theoretical works will help guide the discussion of unconventionality.

## II A. Normal State Experimental Results for $UBe_{13}/U_{1-x}Th_xBe_{13}$

Before discussing the superconducting state, it would be useful to first discuss what is known about the normal state out of which it forms.

Pure $UBe_{13}$, which is cubic (see Fig. 4), has a lattice parameter $a_0$=10.256 Å, while $ThBe_{13}$, which occurs in the same structure, has $a_0$=10.395 Å [5]. Thus, adding Th to $UBe_{13}$ expands the lattice (as did the majority of dopants used by Smith et al. [10] in their study of $U_{1-x}M_xBe_{13}$). Consistent with this, Ott et al. [11] observed a linear-with-x expansion (Vegard's law), with a change in lattice parameter in $U_{0.94}Th_{0.06}Be_{13}$ from the undoped $UBe_{13}$ of about 0.008 Å.

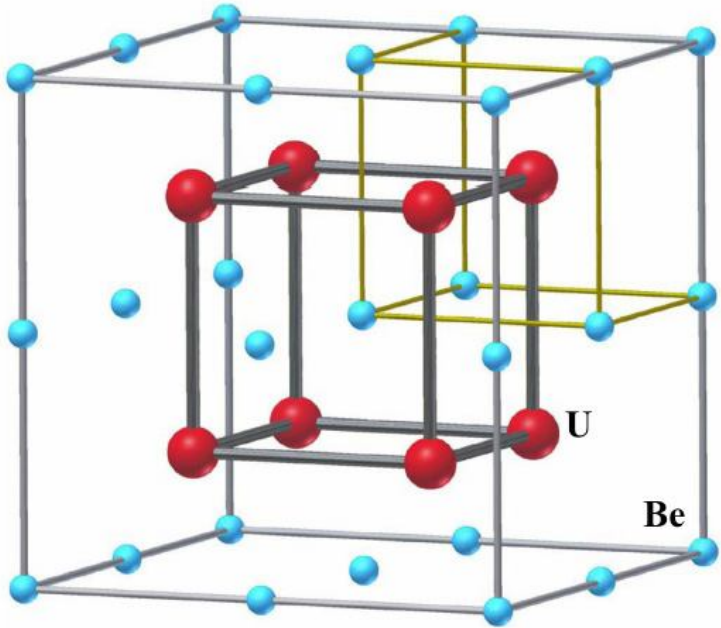

Fig. 4 (color online) Crystal structure of $UBe_{13}$ [13]

All of the heavy fermion, highly correlated 4f- and 5f-electron materials have an enhanced normal state dc magnetic susceptibility (see ref. 6). For $UBe_{13}$, $\chi$ at T≈1 K is [6] 1.5 $10^{-2}$ emu/mole (approximately the same [6] in both single crystal and polycrystalline form). For polycrystalline $U_{0.962}Th_{0.038}Be_{13}$, $\chi$ at 0.7 K is [14] 1.7 $10^{-2}$ emu/mole.

The resistivity, $\rho$, of $U_{1-x}Th_xBe_{13}$ from $T_c$ up to ~3 K has been reported in the discovery work [10] of Smith et al. Undoped $UBe_{13}$ has a peak (of uncertain origin) in $\rho$ vs temperature around 2.5 K (that corresponds to a peak in C – not C/T – vs T, see Figs. 3 and 5), below which $\rho$ falls slightly until the rapid drop off at $T_c$. Below this peak, the temperature dependence of $\rho$ of pure $UBe_{13}$ in field to suppress $T_c$ follows [7] the non-Fermi liquid behavior $T^{3/2}$ from 0.2 to 1 K. For x=0.009, this peak in $\rho$ is [10] shifted down to about 1.2 K (see Fig. 5 for the contrasting C vs T behavior where $T_{peak}$ for x=0.01 is ≈ 1.8 K), and for x=0.026 $\rho$ is flat above $T_c$. Thus, for x≥0.026, the peak in $\rho$ is no longer present.

The normal state specific heat divided by temperature, C/T, of both $UBe_{13}$ (see Fig. 1 above) and $U_{1-x}Th_xBe_{13}$ *increases* with decreasing temperature at low temperatures as T approaches $T_c$ from above. (As will be seen below in the discussion of the superconducting state properties, this increasing C/T as T is lowered has important consequences for the entropy.) Table 1 shows a summary of C/T at 1 K for $U_{1-x}Th_xBe_{13}$.

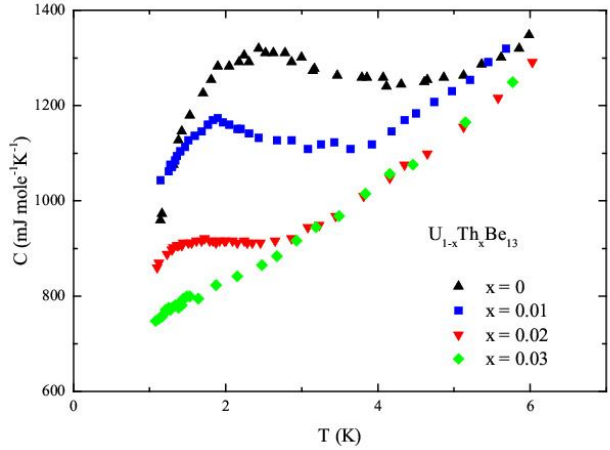

Fig. 5 (color online) Kim et al. [15] Specific heat vs temperature of $U_{1-x}Th_xBe_{13}$ for $0 \leq x \leq 0.03$.

Table 1 Specific heat at 1 K for $U_{1-x}Th_xBe_{13}$ (unannealed polycrystal unless otherwise stated)

| x= | C/T at 1 K in J/moleK$^2$ | reference |
|---|---|---|
| 0 (crystal grown in Al flux) | 1.1 | 8 |
| 0 | 1.0, 0.8 | 5, 11 |
| 0.03 (annealed polyxtal) | ≈0.9 (at 0.75 K) | 12 |
| 0.033 | ≈0.85 | 11 |
| 0.06 | ≈0.9 (at 0.9 K) | 11 |

As is apparent from the table (see also Fig. 3), the few per cent of Th necessary to reach the flat $T_c$ vs x, two-specific-heat-peak region of x ($0.022 \leq x \leq 0.038$) does not appreciably change the normal state value of C/T – proportional to the strongly correlated density of states - just above $T_c$. (As we will see when discussing the superconducting state specific heat below, Th doping does cause a large change in C/T at low (T < 0.3 K) temperatures in the superconducting state.) If we identify the normal state C/T at low temperatures as a metric for the characteristic Fermi temperature $T_F$ of the mass-enhanced, renormalized density of states ($T_F \propto 1/[C/T(1\ K)]$), then $T_F$ (which can serve as an effective band width) is about 20 K. [11] (The basis for this estimate can be found in [6] and involves a simple rigid band model.) Since the Debye temperature (proportional to the lattice stiffness) for the MBe$_{13}$ compounds is [1] about 600-800 K, the result that $T_F << \Theta_D$ is an indication [9,16] of something other than electron-phonon coupling, i. e. is consistent with unconventional superconductivity.

Despite the small size of $T_F$ for UBe$_{13}$ and $U_{1-x}Th_xBe_{13}$, the magnetic field dependence of the normal state specific heat is relatively small. The application [7] of 12 T to UBe$_{13}$ suppresses superconductivity, and results in $C^{normal}/T \propto \log T$ between 0.2 and 3 K, where such a non-Fermi liquid, logarithmic temperature dependence is consistent with the presence of antiferromagnetic spin fluctuations caused by a quantum critical point and also consistent [16] with unconventional superconductivity. This $C/T \propto \log T$ behavior in the normal state with $T_c$ suppressed with field has also been seen [17] in the heavy fermion superconductor CeCoIn$_5$. Interestingly, the normal state thermoelectric power in UBe$_{13}$ also shows [18] a non-Fermi liquid logT behavior in zero field from $T_c$ up to 2 K, which however is quickly suppressed with field.

So a central question in understanding the normal state to be addressed is what parameters influence the formation of this large $\gamma$? One standard (oversimplified) picture often used to explain the formation of the heavy fermion ground state in general is to consider the f-electrons on each ion as Kondo "impurities" and to use the idea of a Kondo resonance to produce the large $\gamma$, large effective mass m* ground state. This is the so-called single-ion Kondo model. Obviously, when each lattice site should have the conduction electrons shielding the local f-electron magnetic moment, the concept of 'impurity' or 'single-ion' should not be applicable. In actuality, the only f-electron, large $\gamma$ systems that the author is aware of where dilution of the f-site leaves $\gamma$ and $\chi$, normalized per mole of f-atom, unchanged as a function of concentration is the antiferromagnet ($T_N$=1.1 K) $Ce_{1-x}La_xPb_3$ for T above 1.5 K, x=0 to 0.8, [19] and $Ce_{1-x}M_xCu_6$, M=La,Th – and in this second case only up to x=0.4. [20].

In the case of diluting $UBe_{13}$ via doping on the U-site, $U_{1-x}M_xBe_{13}$, Kim et al. [5] found that the magnetic susceptibility at 1.8 K for $U_{1-x}M_xBe_{13}$ (for 7 different M) stayed at around 15 memu/mole-U (the value for pure $UBe_{13}$) to ±20% for all dilutions – out to x=0.99 for M=Y. However, for the specific heat $\gamma$ (approximated by Kim et al. as C/T measured at 1.05 K; temperature dependence between 0.35 K and 1.05 K for x>0.05 was negligible) Kim et al. [5] found (see Fig. 6) the following. C/T at 1.05 K ($\approx\gamma$) falls from $\approx$1000 mJ/mole-U-$K^2$ for the undoped compound (C/T as T$\rightarrow$0 is about 1100 mJ/mole-U-$K^2$ in [8]) to an approximately constant 350±50 mJ/mole-U-$K^2$ (approximately independent of x for x$\geq$0.15 as shown in Fig. 6) for dopants M *smaller* than U (M=Hf, Zr, Sc, Lu, Y). For dopants M *larger* than U (M=Ce, La, Th, Pr), $\gamma$ normalized per mole-U was found to fall monotonically with increasing x, never reaching the constant value found for M smaller than U. Thus, in neither case does the total $\gamma$ scale with x as required by the Kondo impurity model (and observed only in $Ce_{1-x}La_xPb_3$ and $Ce_{1-x}M_xCu_6$, M=La, Th, for x $\leq$ 0.4).

Kim et al.'s interpretation of their results was that the separation between the U f-ion and the surrounding Be s- and p-electrons was crucial for the behavior of $\gamma$ with dilution: for $d_{U-Be}$ smaller or equal to that in pure $UBe_{13}$ (i. e. *to the left* of $UBe_{13}$ in Fig. 6), the U f-electrons hybridize with the Be s and p electrons to produce a large, concentration independent (for x$\geq$0.15) $\gamma$ of around 350 mJ/mole-U-$K^2$ – which is a quite sizable value for $\gamma$. (For example, $\gamma$ for pure $UPt_3$ is [6] only 450 mJ/mole-U-$K^2$.) As soon as the $d_{U-Be}$ separation *exceeds* that in pure $UBe_{13}$, there is no concentration independent contribution to the specific heat $\gamma$. Thus, the highly correlated $\gamma$ observed in pure $UBe_{13}$ has two parts. 1.) About 65% (or 650 mJ/mole-U-$K^2$) is due to correlation effects (i. e. not single ion behavior) between the U f-ions that start to disappear for dilution already much less than 15%. For example, $\gamma$ for $U_{0.97}Y_{0.03}Be_{13}$, $a_0$=10.2385 Å, is [5] only 620 mJ/mole-U-$K^2$ (not shown in Fig. 6.) 2.) About 35% (or 350 mJ/mole-U-$K^2$) comes from hybridization effects that extend to the very dilute limit (i. e. do exhibit scaling with x and no coherence between the U 5f ions is required) for $d_{U-Be}$ less than or equal to the value in pure $UBe_{13}$. For dopants that increase $d_{U-Be}$, $\gamma$ falls to low, normal metallic values monotonically with increasing doping and has no concentration independent part. (As an aside, these dilution results – as shown by the immediate destruction of the large coherent value of $\gamma$ in, e. g., $U_{0.97}Y_{0.03}Be_{13}$ with a change in lattice parameter of only 0.54 $10^{-3}$ Å - of Kim et al. are not some equivalent form of 'chemical pressure' (where the dopant is either larger, equivalent to negative pressure, or smaller, equivalent to positive applied pressure) comparable to pressure measurements of γ in concentrated $UBe_{13}$. Specific heat under pressure of pure $UBe_{13}$ by Olsen et al. [21] found a 30% decrease in γ at 0.89 GPa. Using the known [22] bulk modulus of $UBe_{13}$ of 108 GPa, this 30% change in γ under pressure in pure $UBe_{13}$ (comparable to that found by doping 3% Y in $U_{0.97}Y_{0.03}Be_{13}$) corresponds to a change in lattice parameter of 28.5 $10^{-3}$ Å.)

Thus, since there is no model to explain this rather diverse behavior of γ in $UBe_{13}$ and its connection to the quantum critical logT behavior of the specific heat, the goal of trying to understand the ground state out of which the superconducting state forms is still an open question.

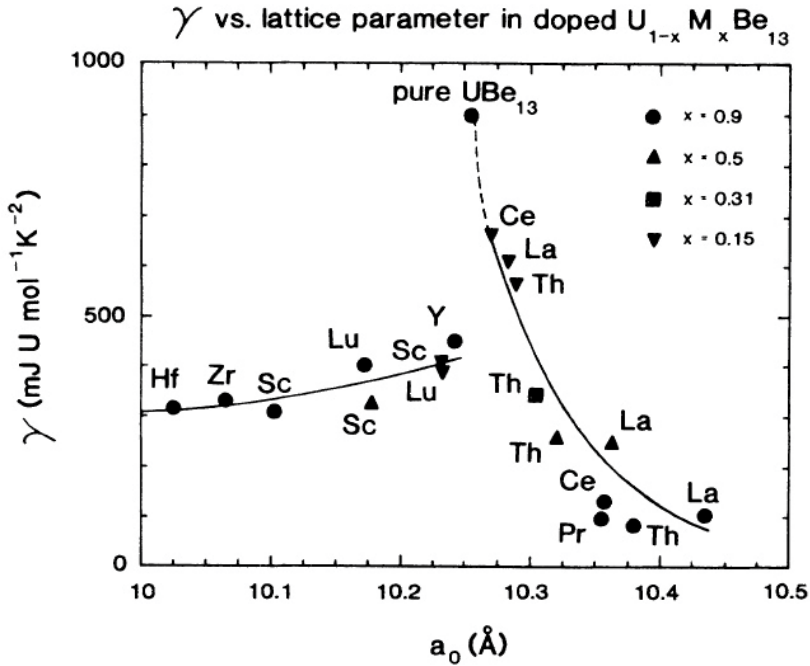

Fig. 6 Specific heat γ vs lattice parameter in $U_{1-x}M_xBe_{13}$ (lattice parameter of pure $UBe_{13}$ is 10.256 Å) from Kim et al. [5] showing that correlation effects are important for the heavy fermion ground state in this compound, with partial scaling behavior evident for dilutants smaller than U.

## II B.  Superconducting State Experimental Results for $UBe_{13}$/$U_{1-x}Th_xBe_{13}$
### 1.  Specific Heat/Ultrasonic Attenuation/Lower Critical Field - Overview

There is an enormous amount of characterization that has been carried out on $UBe_{13}$, discovered to be a superconductor in 1983, and on $U_{1-x}Th_xBe_{13}$, discovered to have *two* superconducting transitions – clear proof of unconventional superconductivity – shortly thereafter. The discussion below is organized more or less according to the timeframe when the work was reported, with some discussion of results (e. g. the specific heat discussed just below) out of time sequence when such discussion is best considered together with earlier measurements which directly led to the later discovery.

Overhauser and Appel, in 1985, pointed out [23] that the superconducting specific heat in $UBe_{13}$ can be scaled onto that of the known BCS, electron phonon coupled superconductor Pb. Thus, they argue that pure $UBe_{13}$ is a conventional, BCS superconductor. As we will see in the discussion below, this initial observation does not match with the consensus today.

Concerning $U_{1-x}Th_xBe_{13}$, as we saw in the previous section, the normal state susceptibility and specific heat of $U_{1-x}Th_xBe_{13}$ measured above $T_c$ do not change appreciably in the range $0 \leq x \leq 0.06$. This is <u>not</u> the case for the specific heat below $T_c$.

Several sets of measurements of the bulk specific heat have been performed to determine, among other properties, the phase diagram of $T_c$ (both the upper transition, $T_{c1}$, and the lower, $T_{c2}$) vs x in $U_{1-x}Th_xBe_{13}$ in the region where Smith et al. [10] (Fig. 2) saw $T_c$ vs x flatten out. Since, as will be discussed below, at least one if not both of the transitions in $U_{1-x}Th_xBe_{13}$ are caused by unconventional superconductivity (which is known [16] to be sensitive to *both* magnetic and non-magnetic impurities), the $T_c$ vs x results may well be dependent on sample quality. Thus, we will use the phase diagram of Scheidt et al. (see Fig. 7) based on unannealed *high purity* polycrystalline specimens made from electrotransport refined U from Ames Laboratory and "MBE" grade 99.999% pure Be from Atomergic. Such high purity U generally has very low impurities, e. g., approximately 11 atomic ppm Fe. We will also discuss the comparison of Scheidt et al.'s phase diagram to other phase diagrams based on the use of normal purity U (40-180 atomic ppm Fe) and 99.5 % Be (from Brush Wellman, approximately 500-1000 atomic ppm Fe). As well, when the nature of the two transitions is discussed, the effect of long term annealing of the high purity samples will be considered.

Note from the phase diagram (Fig. 7) for the two transitions seen in unannealed high purity $U_{1-x}Th_xBe_{13}$ that $T_{c2}$ for x just above the left hand critical concentration, $x_{c1}$, for two transitions falls markedly with increasing x, reaches a minimum at x=0.03, and then rises until $x_{c2}$ is reached. In contrast, the phase diagram of Ott [25] based on normal purity material has $T_{c2}$ approximately constant between $x_{c1}$ and $x_{c2}$. Also, note that the phase diagram in Fig. 7 – using specific heat data (see Figs. 8 and 9 below) from high purity samples – has the region where $T_{c1}$ is approximately constant extend past 4 %. This is in contrast to the $\chi_{ac}$ data of Smith et al. [10] in Figure 2 showing a flat $T_{c1}$ vs x region extending only up to 3%. However, the specific heat data on regular purity $U_{1-x}Th_xBe_{13}$ of Ott et al. [11] have two 'endpoint' compositions, x=0.0216 and 0.0378, whose $T_{c1}$ and $T_{c2}$ values (but not the $\Delta C/T_{c1}$ values – see Figs. 8 and 9) agree well with the Fig. 7 diagram here for x=0.022 and 0.038. Thus, at least the intercomparability/reproducibility of x for the values of measured $T_{c1}$ and $T_{c2}$ of $U_{1-x}Th_xBe_{13}$ between different laboratories and materials purities seems quite good.

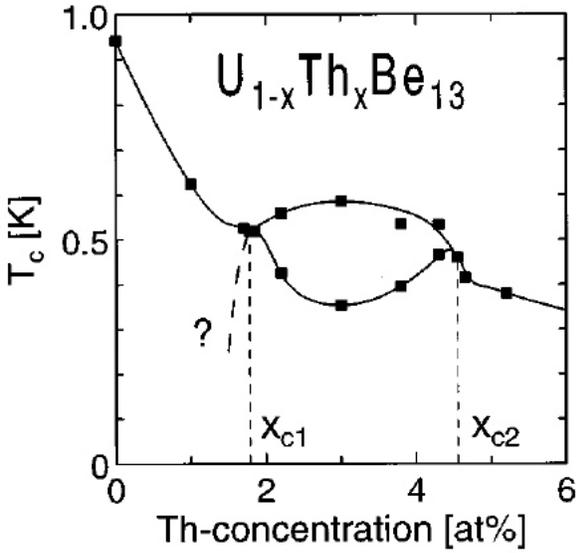

Fig. 7: Superconducting transition temperature $T_c$ in high purity $U_{1-x}Th_xBe_{13}$, after Scheidt et al. [24] The question of a second anomaly in the specific heat for x<0.018 (marked by a dashed line and a question mark in the figure) will be discussed below. The compositions $x_{c1}$ and $x_{c2}$ marked by the vertical dotted lines are the boundaries for the two phase region.

The specific heat data [24] on which the phase diagram of Fig. 7 is based are shown in Figs. 8 and 9. Note the rather sudden appearance of a second transition in the specific heat at x=0.022 (i. e. there is no second transition at x=0.0185), and further note that at the x=0.022 composition (as shown in the phase diagram in Fig. 7), $T_{c1}$ actually *increases* above $T_c$ for x=0.0185, in contrast to the phase diagram in Fig. 2 of Smith et al. [10]. Note also that $\Delta C/T_c$ for x before the two transition composition at 2.2% Th (see Fig. 8) has a very unusual trend as x increases: at x=0.01 and 0.017, $\Delta C/T_c$ remains ~constant, while $\Delta C/T_c$ jumps by 70% from x=0.017 to 0.0178. This increased $\Delta C/T_c$ then remains constant as x increases to 0.0185, and then splits into two transitions at x=0.022.

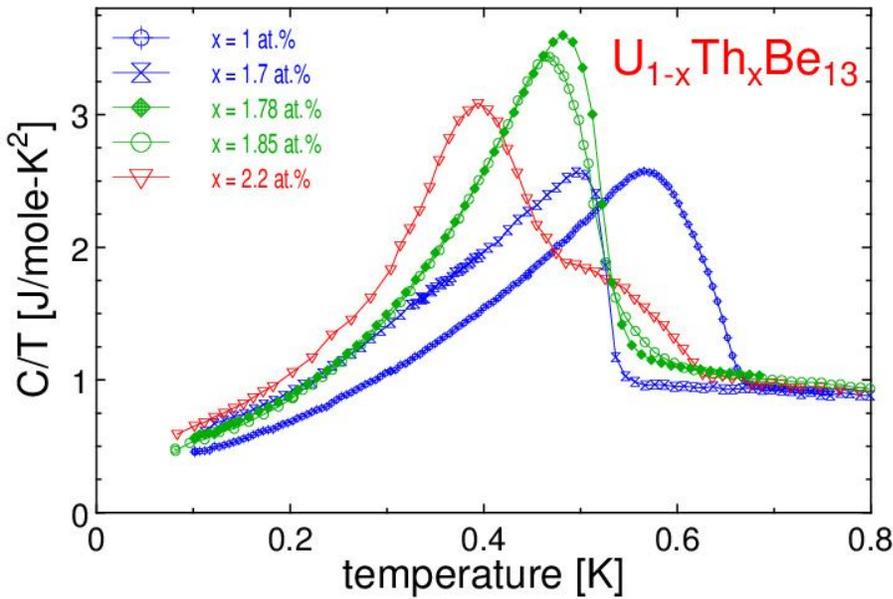

Fig. 8. (color online) Specific heat divided by temperature, C/T, vs temperature, T, from Scheidt, et al. [24] for 0.01≤x≤0.022.

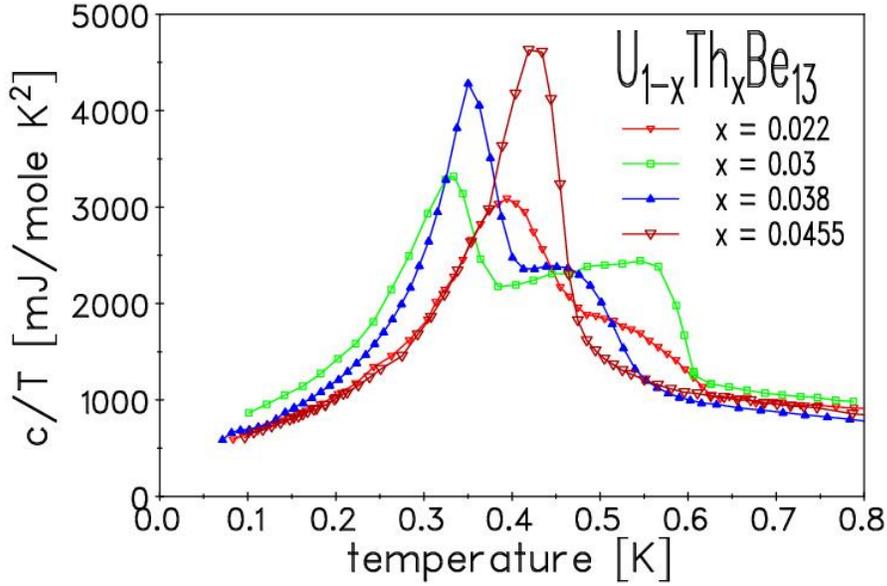

Fig. 9. (color online) Specific heat divided by temperature, C/T, vs temperature, T, from Schreiner, et al. [26] for 0.022≤x≤0.0455. Note that the discontinuity in the specific heat divided by temperature, $\Delta C/T_{c1}$, at the upper transition for x=0.03 (which, as clear from these data as well as from the phase diagram in Fig. 7, is in the middle of the two transition region) is significantly narrower than that for x=0.022.

In the discovery paper [11] of Ott et al. of a bulk anomaly in the specific heat at both $T_{c1}$ and $T_{c2}$, the nature of the lower transition at $T_{c2}$ was uncertain. Various possibilities were discussed, with a magnetic transition judged to be unlikely since NMR measurements [27] (actually published before the discovery paper [11] because of the vagaries of the review process) on $U_{1-x}Th_xBe_{13}$, x=0.03, found no sign of magnetic order (with an error bar for the ordered moment of about 0.01 $\mu_B$) at $T_{c2}$.

One way to approach trying to understand the nature of the two superconducting transitions in $U_{1-x}Th_xBe_{13}$ is alluded to in the discussion above of the sharp increase in $\Delta C/T_c$ between x=0.017 and 0.0178. Namely, one should consider the evolution of the entropy associated with the two transitions in the superconducting state as a function of x. This analysis was performed by Schreiner et al. [26] and is shown here in Fig. 10.

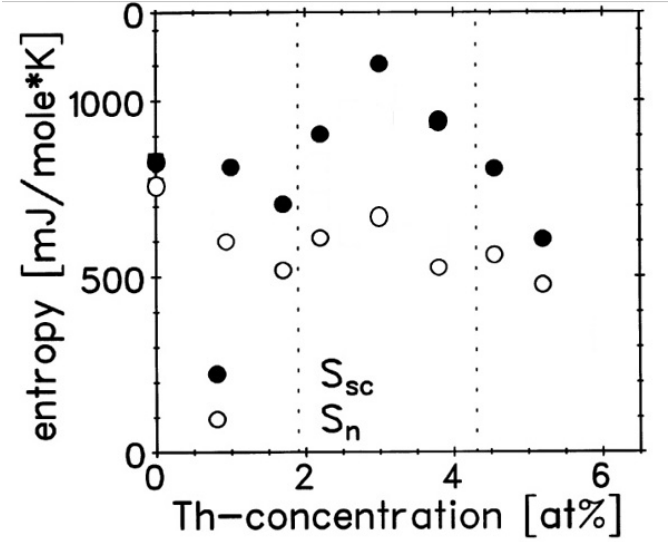

Fig. 10: Entropy [26] for $U_{1-x}Th_xBe_{13}$ *measured* in the superconducting state up to $T_c$ ($S_{sc}$) and normal state entropy at $T_c$ ($S_n$) obtained by *assuming* that C/T in the extrapolated normal state is constant below $T_c$, i. e. where the integral for $S_n$ between 0 and $T_c$ is given by $S_n \equiv \int (C/T)|_{T_c} dT$. Obviously, since C/T measured in the normal state above $T_c$ is *increasing* with decreasing temperature (see Fig. 9) for the compositions between the dotted vertical lines (compositions labeled $x_{c1}$ and $x_{c2}$ in Fig. 7), making this assumption here is a method to visually emphasize how strongly $C/T_{normal}$ in applied fields below $T_c(H=0)$ must increase to match the measured superconducting state entropy at $T_c$, $S_{sc}$.

Clearly, in order for the entropies to match at $T_c$ as required for a second order phase transition (there is no evidence that either of the transitions in $U_{1-x}Th_xBe_{13}$ is first order), the normal state C/T values extrapolated below $T_c$ for each x > 0 composition shown in Fig. 10 must *increase* below the superconducting transition. This increase upon lowering temperature has already been remarked on above as one indication of non-Fermi liquid behavior in these materials. Certainly a further increase of $C/T_{normal}$ below $T_c$ (either extrapolated or measured in a magnetic field to suppress superconductivity) is consistent with the trend in $C/T_{normal}$ shown in Fig. 9. As shown in Fig. 10, the $S_{sc}$ measurements in $U_{1-x}Th_xBe_{13}$ imply that the degree of non-Fermi liquid behavior ($\Leftrightarrow$ the rate of rise of $C/T_{normal}$ as $T \rightarrow 0$) is maximum at x=0.03 – i. e. in the middle of the two phase regime.

This peaking of non-Fermi liquid behavior at x=0.03 implies a connection between the two phase transitions in $U_{1-x}Th_xBe_{13}$ between $x_{c1}$ and $x_{c2}$ (and the associated maximum in the difference between the entropies $S_{sc}$ and $S_n$) with a strong influence on the temperature dependent renormalized effective mass m* caused by doping U with Th. How Th affects this non-Fermi liquid behavior (temperature dependence of $C/T_{normal}$) in $U_{1-x}Th_xBe_{13}$ in this limited composition range is as yet unanswered and remains a challenging problem for theorists.

Following this (still only partial) discussion of the specific heat of the two transitions in $U_{1-x}Th_xBe_{13}$, we return now to the timeline of discovery of the properties of the two transitions in $U_{1-x}Th_xBe_{13}$, after Ott et al.'s discovery [11] of two specific heat transitions.

Batlogg et al. [28] made the next contribution to the discussion of the nature of the lower transition by reporting a very large peak in the ultrasonic attenuation in $U_{1-x}Th_xBe_{13}$, x=0.0175, at 0.39 K and an onset of the transition at 0.45 K, where $T_c$ onset from the magnetic susceptibility is about 0.6 K, with the susceptibility transition complete at just above 0.4 K. This sample was made in the same laboratory as those in the discovery works [10-11]. Batlogg et al. then inferred that the ultrasonic transition at lower temperature was due to antiferromagnetism. Five years later, μSR work [29] by Heffner et al. did find a moment "of order" 0.001-0.01 $\mu_B$ beginning below $T_{c2}$ for x=0.0193, 0.0245, and 0.0355 – i. e. only in the region of Th-concentration where there are two anomalies in the specific heat. Note that this moment is too small to explain the entropy associated with the specific heat transition at $T_{c2}$. From the current perspective it should be pointed out that intervening specific heat work [24] on high quality samples with a very precise variation in x, with values of x equal to 0.0170, 0.0178, and 0.0185 (see Fig. 8), saw *no evidence* of a second transition in the specific heat at or near x=0.0175. (The original work [11] by Ott et al. reported that the lowest concentration where a second peak in C/T occurred was x=0.0216. A succeeding work [30] by the same group, also using normal purity U and Be, reported that the occurrence of a second phase transition is "first observed for x=0.0205." A study of the increase in $|dH_{c1}/dT|$ at the lower transition in $U_{1-x}Th_xBe_{13}$ by Knetsch et al. [31] found evidence for a second transition at x=0.0193.)   Thus, the Batlogg et al. discussion of a characteristic of the lower transition at $T_{c2}$ being a large peak in the ultrasonic attenuation is suspect since their stated concentration of Th was not in the range $x_{c1}<x<x_{c2}$ but was in fact *below* the concentration where the onset of a second transition is first observed.

The next, and fairly definitive, characterization of the lower transition was when Rauchschwalbe et al. found [32] that the absolute magnitude of the rate of change with temperature of the lower critical field, $H_{c1}$, (or $–dH_{c1}/dT$) increased by more than a factor of three as temperature was decreased below $T_{c2}$. This finding is consistent with a significant increase in the density of superconducting quasiparticles below $T_{c2}$. They proposed that a second portion of the Fermi surface becomes superconducting below $T_{c2}$.

Before discussing the various theories proposed to explain this puzzling phenomenon, let us consider some of the characterization results in greater depth.

## 2. Anomalies below $T_{c1}$ in pure $UBe_{13}$, x=0; Tracking of a possible second anomaly below $T_{c1}$ for x<$x_{c1}$:

Finding an extension into the region of x<$x_{c1}$ of the two phase lines which join at a point (Fig. 7) as x → $x_{c1}$ from above has been a major unanswered question in the study of $U_{1-x}Th_xBe_{13}$.

### x=0:

1.) There is a report by Rauchschwalbe et al. [32-33] of a broad second anomaly in the specific heat *as a function of temperature* of pure $UBe_{13}$. This possibility is of some interest, because then the phase diagram shown in Fig. 7 would have a second line of transitions to the left of $x_{c1}$ below the single line shown. This lower line (which would have $T_c$ change slowly with x, i. e. not coincide with the dashed line in Fig. 7 marked by '?') would join/lead to one of the two $T_c$ vs x lines between $x_{c1}$ and $x_{c2}$, making for a more continuous transition at the left hand critical concentration, and satisfying thermodynamic constraints as discussed below.

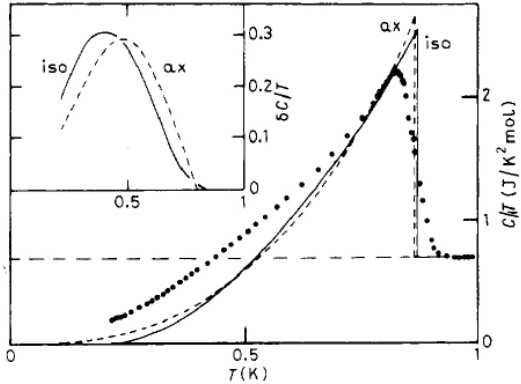

Fig. 11 Specific heat divided by temperature [32-33] of pure $UBe_{13}$. The solid and dashed lines are fits to theory (isotropic and axial symmetry respectively). [32-33] When the difference between the two theoretical fits and the data are plotted in the inset, a definite anomaly is visible at about 0.5 K. The original data shown here, and analyzed as discussed by Rauchschwalbe et al. [32-33], is from Mayer et al. [34] on an arc-melted polycrystalline sample of normal purity.

Rauchschwalbe et al. [32-33] found an indication of a second transition (see Fig. 11) at around 0.5 K in zero field in the specific heat of *pure* $UBe_{13}$, although the sample quality of the polycrystalline sample was not optimal: the peak in C/T was at 0.8 K with $T_c^{onset}$ = 0.92 K (i. e. width of the superconducting transition $\Delta T_c$=0.12 K), vs the values [12] for the annealed, high purity sample whose data were shown above in Fig. 3 of $T_{peak}$=0.964 K and $T_c^{onset}$ = 0.992 K ($\Delta T_c$=0.028). They [33] then joined this point to the *upper* $T_c$ vs x line in the two-transition region in Fig. 7.

2.) Differing from Rauchschwalbe et al. [33], Ellman et al. [35] in their specific heat data <u>as a function of *magnetic field*</u> found clear evidence of a rather distinct second transition in high purity (electrotransport refined U and 99.999 % pure Be) polycrystalline $UBe_{13}$. Their data and the accompanying phase diagram are shown in Figs. 12 and 13. The two sets of data, from Rauchscwalbe et al. [32-33] in zero field and from Ellman et al. [35] in applied field, cannot apparently be reconciled with each other. Ellman et al.'s result has been confirmed by Kromer et al. [36], who measured a single crystal of $UBe_{13}$, with $T_c$=0.90 K, $\Delta T_c$=0.025 K. Kromer et al. see an anomaly in their C vs H data also at T=0.63 K at about 1.5 T, which would extend the lower phase line of Ellman et al. in Fig. 13.

Fig. 12 Specific heat vs temperature [35] of high purity $UBe_{13}$ at different fixed temperatures as a function of field. Note the distinct anomalies (e. g. in 0.4 and 0.5 K)

Fig. 13 Phase diagram in the H-T plane for the observed two anomalies in Fig. 12 [35]. The lower anomaly (here vs field) is inconsistent with the anomaly in

*below* the higher temperature transition.

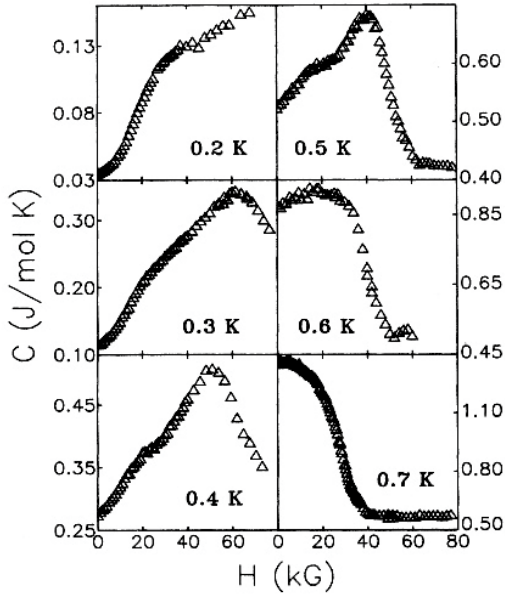

zero field at 0.5 K of Rauchschwalbe et al. [33]. Note the large critical field slope, -$dH_{c2}/dT$, at $T_c$ for pure $UBe_{13}$ – the largest of any heavy Fermion superconductor [6]

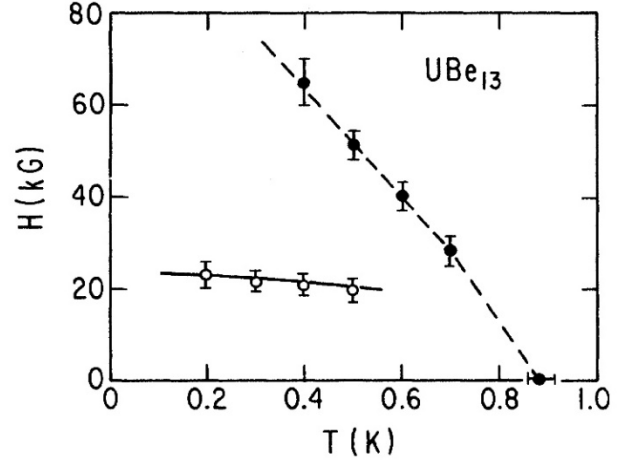

3. Recently, Shimizu et al. measured [37] the specific heat of a single crystal (grown by the Al flux method) of $UBe_{13}$ to temperatures down to 0.075 K and in fields up to 5 T. They found no evidence of nodal behavior, both in their angular $C(H, T, \Theta)$ results and in the field dependence ($C/T \propto H$) at 0.08 K. (Nodal behavior would result [38] in $C/T \propto H^{1/2}$ at temperatures well below $T_c$.) In addition, Shimizu et al. [37] found an anomaly (a 'weak hump' in $C/T$ above 3 T) consistent with the data of Ellman et al. [35] shown in Fig. 12 and of Kromer et al. [36]

**x very near to $x_{c1}$:** By measuring specific heat under uniaxial stress of a single composition (x=0.022) of high purity $U_{1-x}Th_xBe_{13}$, and by using a very small minimum step size of 0.1 kbar ($10^7$ Pa), Zieve et al. [39] were able to scan very precisely the phase diagram in Fig. 7 near $x_{c1}$. They did indeed find a transition like that represented by the dashed line (i. e. almost vertical and parallel to the T-axis) in Fig. 7. Their data, see Fig. 14, supply important constraints for understanding the multiple (presumably unconventional) superconducting states in $U_{1-x}Th_xBe_{13}$ and the joining of the various phase transition lines at the polycritical point at $x_{c1}$ (discussed below).

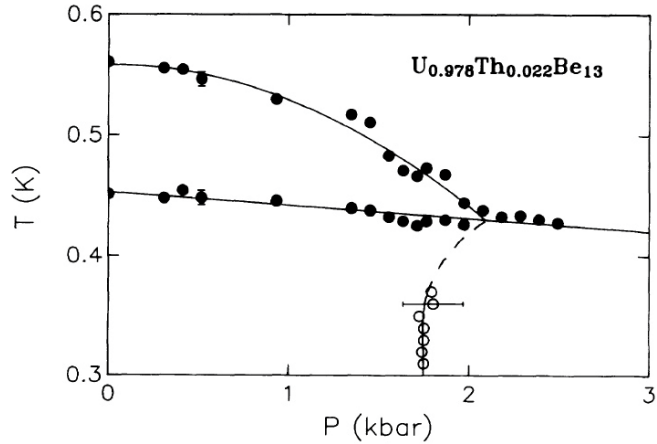

Fig. 14. Zieve et al. [39] phase diagram derived from specific heat under uniaxial pressure. Note how the two specific heat transitions ΔC for x=0.022 join for P≥2 kbar, as well as the appearance of a new, third transition, just to the left of this critical pressure.

**x below $x_{c1}$:** Expanding on the phase diagrams of Figs. 7 and 13, Lang et al. [40] found a clear, separate (from the upper transition at $T_{c1}$) transition in the thermal expansion of high quality polycrystalline $U_{1-x}Th_xBe_{13}$ for x=0.03 and followed this anomaly to lower x (where it became smeared together with the drop in the thermal expansion at $T_{c1}$ and, for x<$x_{c1}$, with the drop in the thermal expansion at $T_c$), proposing a line of $T_L$ vs x for x<$x_{c1}$ (see Fig. 15) that is consistent with the Rauchschwalbe et al. proposal that there is a second transition below the $T_c$ vs x line in Fig. 7 for x<$x_{c1}$. (This second transition starts for x=0 around 0.75 K for Lang, et al. and 0.5 K for Rauchscwhalbe et al., i. e. not the sharp fall off almost parallel to the T-axis as measured by Zieve et al., which was consistent with the dashed line and '?' in Fig. 7.) Rather than the Sigrist and Rice [41] proposition for different superconducting states (discussed thoroughly below),

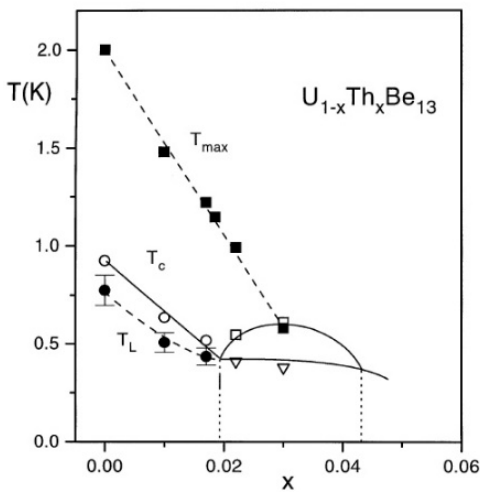

Fig. 15. Phase diagram for $U_{1-x}Th_xBe_{13}$ from Lang et al. [40] The three solid circles indicate a proposed phase transition from features in the thermal expansion, α. The solid squares refer to maxima in the *normal state* α vs temperature data, and are roughly equivalent to the peaks in the specific heat vs temperature (Fig. 5), e. g. $T_{peak}$ in C vs T from Fig. 5 for x=0.01/0.02 is 1.8/1.5 K vs 1.5/1.2 K for Lang et al.'s α vs T data for x=0.01/0.022.

Lang et al. argue that the lower transition: a.) for x<$x_{c1}$ is due to short range antiferromagnetic correlations and b.) for $x_{c1} \le x \le x_{c2}$ is due to long range antiferromagnetic order.

In summary, anomalies in the specific heat and thermal expansion in pure $UBe_{13}$ at temperatures (0.5 K Rauchschwalbe et al., Lang et al. 0.75 K) not unlike that of the second, lower superconducting transition in $U_{1-x}Th_xBe_{13}$ remain a puzzle that is further addressed below in the theory section.

### 3. Specific heat of $U_{1-x}Th_xBe_{13}$ for $0.090 \leq T \leq T_c$ for the two phase region $x_{c1} \leq x \leq x_{c2}$:

Already above with Figs. 7-9, these data were briefly overviewed, and are discussed in more depth here. Returning to the question of the nature of the lower transition in $U_{1-x}Th_xBe_{13}$, Kim et al. [12] performed long term annealing on an x=0.03 sample made from high purity U and Be. If we compare these results with those of Schreiner et al. [26] (Fig. 9) on an unannealed piece of the same sample (prepared at the University of Florida), we find the result shown in Fig. 16. The value for $\Delta C/T_{c2}$ more than doubles upon annealing; as well, the width of the superconducting transition in the unannealed sample is reduced. Annealing is thought to remove non-magnetic impurities. Thus, this comparison of the specific heats of annealed [12] and unannealed [26] high purity $U_{0.97}Th_{0.03}Be_{13}$ is consistent with the transition in $U_{1-x}Th_xBe_{13}$, x=0.03, at $T_{c2}$ being unconventional. However, as can be seen in Fig. 16, the upper transition at $T_{c1}$ is not affected to such a marked degree by annealing, which would be consistent with the upper transition being due to conventional superconductivity.

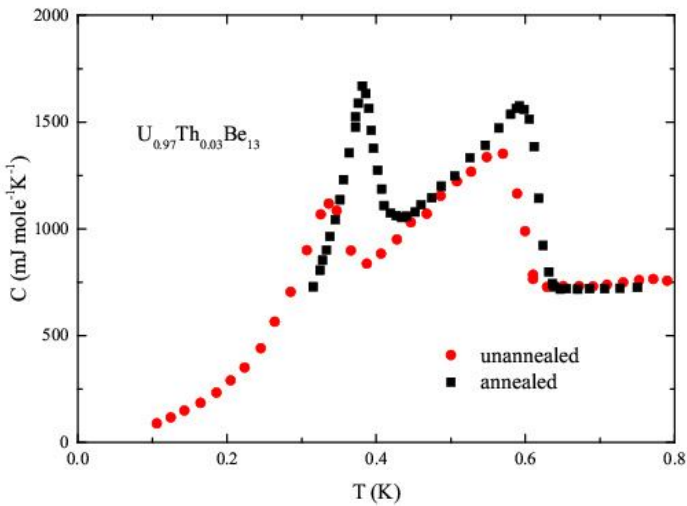

Fig. 16 (color online) Kim et al. [12] data are black squares (high purity sample annealed at 1400 C for 7.3 weeks) and Schreiner et al. [26] unannealed data on the same high purity sample (red circles). The lower transition is much more sensitive to defects removed by annealing than the upper transition.

Another method of comparing the nature of the upper and lower transitions in the two phase region $x_{c1} \leq x \leq x_{c2}$ would be to consider their rate of suppression with an applied magnetic field. An initial measurement [42] of the specific heat in magnetic field of x=0.0331 by Ott et al. found that above 1 T the rate of shift downwards of $T_{c2}$ with field was somewhat smaller than for $T_{c1}$. This is consistent with the work [43] of Jin et al. who, using high purity material, also found that above 1 T the two transitions in x=0.03 as a function of field came together, extrapolating to a zero temperature critical field for both transitions of about 4.8 T. (This can be compared to $H_{c2}$(T=0) of pure $UBe_{13}$ of between 11 [44] to 14 [45] T.) However, in a measurement with five finite field measurements up to 1.25 T of $T_{c1}$ and $T_{c2}$ in high purity material, Kim et al. [12] found that the two transitions – at least in this low field limit – were suppressed at the *same* rate with magnetic field. Their conclusion was that the two transitions were of rather similar nature.

Another investigation of the specific heat of $U_{1-x}Th_xBe_{13}$ in the two transition region was the work [46] of Jin et al. to lower (0.09 K) temperatures. One of the goals of that work was to be able to discuss if there is an identifiable power law in the superconducting state specific heat for $U_{1-x}Th_xBe_{13}$, where $C \propto T^2$ implies line nodes in the superconducting gap and $C \propto T^3$ implies point nodes. Of course, in order to be significant such power laws should if possible be measured over a decade in temperature. Measurements were made on the compositions shown in Table 2, where all the polycrystalline samples were made from high purity U and Be. For x=0, 0.022, and 0.03, samples were annealed at 1400 C for long periods (3.6, 5.7, and 7.3 weeks respectively.) (Vapor loss of Be during annealing was prevented by including elemental Be as a vapor source in each of the sealed Ta annealing vessels.) The single crystal with x=0 grown from Al flux was annealed at 1100 C for six months. Single crystals of $UBe_{13}$ made using the Al flux method have recently been characterized [47] to contain approximately 0.8 wt. % Al, or $UBe_{12.9}Al_{0.1}$. This unintentional doping is at least a partial reason for the degradation in properties (lower $T_c$, higher residual $\gamma_0$) between the single crystal and the polycrystalline sample of undoped $UBe_{13}$ shown in Table 2. A more thorough characterization by Amon et al. [48] of the effect of *intentionally* added Al in polycrystalline $UBe_{13}$ found that Al is soluble in $UBe_{13}$, and substitutes for Be, up to approximately x=0.3 (i. e. up to $UBe_{12.7}Al_{0.3}$). A further aspect of this work is a study [48] of the annealing effects at 900 °C up to 2 ½ months of the specific heat of single crystals grown in Al flux.

In addition to determining power laws (discussed below), interestingly, Jin et al. discovered large finite C/T values (see Table 2) in the superconducting state when T was extrapolated to zero in their work. For x=0.022 and 0.052, Table 2 shows that these low temperature C/T (T→0) (called $\gamma_0$ in Table 2 and ref. 46) values are comparable to the value of $C^{normal}$/T for T just above $T_c$, see Fig. 9. (This result of large $\gamma_0$, in data for normal purity samples of $U_{1-x}Th_xBe_{13}$, was also qualitatively noted "for some" compositions by Felder et al. [30]) For pure $UBe_{13}$, in addition to the values for $\gamma_0$ reported by Jin et al. listed in Table 2, Ravex et al. [49] reported $\gamma_0$=0.11 J/molK$^2$ for a $T_c$=0.845 K normal purity sample. For a fully gapped superconductor, of course, C/T in the superconducting state should extrapolate to zero. The presence of these large finite $\gamma_0$ values could have various explanations, some of which (discussed below in the theory section) would be consistent with unconventional superconductivity. Since the polycrystalline samples in ref. 46 were made from high purity U and Be, and three of the samples were long-term, high temperature annealed, presumably large amounts of inhomogeneous normal regions/defects are unlikely.  Also, the fact that ref. 30 saw qualitatively similar large $\gamma_0$ in samples of normal purity without annealing argues for an intrinsic explanation.

Table 2: Specific heat study of polycrystalline (except for the annealed x=0 single crystal) $U_{1-x}Th_xBe_{13}$ by Jin et al. [46].  The values derived for $2\Delta/k_BT_c$, when compared to the weak coupling BCS value of 3.52, imply strong coupling.

| x | $T_c$ (mK) | $T_{c2}$ (mK) | $\gamma_0$(J/molK$^2$) | $2\Delta/k_BT_c$ |
|---|---|---|---|---|
| 0 (annealed single xtal) | 860 |  | 0.19 |  |
| 0 (annealed) | 920 |  | 0.07 |  |
| 0.017 (unannealed) | 520 |  | 0.59 |  |
| 0.022 (annealed) | 560 | 465 | 1.06 | 5.2 |
| 0.030 (annealed) | 620 | 395 | 0.75 | 5.6 |
| 0.052 (unannealed) | 375 |  | 1.21 | 5.4 |

Regarding fits to the low temperature C data to search for the presence of power laws, Jin et al. [46] found that for $x<x_{c1}$ (i. e. for x=0 and 0.017) $C/T = \gamma_0 + \beta T^n$, where n=2±0.2.  This power law for C ($\propto T^3$) is consistent with an axial superconducting state with point nodes, and has been observed (although without the inclusion of a large $\gamma_0$) before [50]. For the remaining three compositions of $U_{1-x}Th_xBe_{13}$ measured by Jin et al. (x=0.022, 0.030, and 0.052) the best fit to the superconducting state data below 0.8 $T_c$ (or $T_{c2}$ in the case of x=0.022 and 0.030) is $C/T = \gamma_0 + \beta T^n$, where n=3±0.2.  Since $C \propto T^4$ does not fit any theoretical prediction, Jin et al. called it 'unphysical' and instead fit the data to $C/T = \gamma_0 + \alpha C_{BCS}/T$, where $\alpha$ is just a constant and the values for $C_{BCS}$ come from numerical evaluation of the integrals.  In order to better fit the data, Jin et al. allowed the energy gap $\Delta$ to float, and the results for $2\Delta/k_BT_c$ from the fits are shown in Table 2.  Jin et al. comment that the BCS temperature dependence gave convincingly good fits to the data.  Obviously, it is a bit unusual to have a fit assuming a fully gapped Fermi surface (the $\alpha C_{BCS}$ term) coincident with a large residual $\gamma_0$.

Recent tour-de-force C(T, H, $\Theta$) data [51] on a single crystal (mechanically detached from a tri-arc melted sample, i. e. no Al [47] contamination) suggest that the superconducting gap is fully open over the Fermi surface in $U_{0.97}Th_{0.03}Be_{13}$ below $T_{c2}$ (ignoring the large $\gamma_0$, consistent with the fit [46] of Jin et al. to the temperature dependence of their specific heat data). Since such C(T, H, $\Theta$) measurements can be quite challenging, there still exists the possibility that oscillations of C/T with angle (consistent with nodal behavior) of *smaller magnitude*, could be resolved with background noise improved over that (~3% at 0.08 K and 1 T) achieved in ref. 51.  However, in agreement with their C(T, H, $\Theta$) data, Shimizu et al. [51] also found that instead of $C/T \propto H^{1/2}$, as found [38] for nodal behavior, the low temperature C/T at 0.08 K of their $U_{0.97}Th_{0.03}Be_{13}$ crystal is linear [51] with H up to 1 T, consistent with a fully open gap.  Several gap symmetries can be derived [51] from these results.  The authors also find [51] the absence (presence) of anisotropy for $T_{c2}$ ($T_{c1}$) with field, which they interpret as distinguishing the gap symmetry in the $T < T_{c2}$ superconducting phase from that of the phase between $T_{c2}$ and $T_{c1}$.

In addition to the work of Jin et al. [46] in trying to determine the nature of the superconductivity in $U_{1-x}Th_xBe_{13}$ via observation of power laws in the low temperature specific heat, various authors with the same goal have measured the temperature dependence of other microscopic properties.

Ravex et al. [49] report a large term linear in the temperature of the thermal conductivity, $\kappa$, for their $UBe_{13}$ sample below 0.100 K, but assign its origin to non-intrinsic causes. Jaccard et al. [52], in $\kappa$ data down to 0.13 K in a $T_c$=0.854 K sample of $UBe_{13}$, report a $\kappa \sim T^2$ behavior up to about 0.4 K. A priori such a temperature dependence would be consistent with unconventional superconductivity with polar pairing, except that theory [53] argues against such a pairing symmetry unless [54] there is spin anisotropy in the pairing interaction. Additionally, a power law over such a restricted temperature range must be viewed as only indicative. More recent work by K. Izawa (to be published) looked at the superconducting gap structure in $U_{1-x}Th_xBe_{13}$ using thermal conductivity.

The temperature dependence of the spin lattice relaxation rate, $1/T_1$, in NMR measurements of regular purity $UBe_{13}$, $T_c$=0.78 K, behaves [55] as $T^3$ from $T_c$ down to 0.2 K (i. e. again a limited temperature range). The authors described this as consistent with unconventional superconductivity with a triplet p-wave polar state with line nodes in the gap function. µSR measurements on $U_{0.965}Th_{0.035}Be_{13}$ to determine the spin susceptibility down to 0.02 K were analyzed [56] as consistent with (but not conclusive evidence for) odd parity spin triplet superconductivity below $T_{c2}$=0.35 K.

### III.     Summary of the Experimental Situation:

So, the fundamental question to be answered is: what is the likely pairing symmetry of undoped $UBe_{13}$ and $U_{1-x}Th_xBe_{13}$, both for x in the two phase regime and for $x \leq x_{c1}$? Is the superconductivity in pure $UBe_{13}$, with its low residual $\gamma_0$ in the superconducting state, different from the superconductivity occurring at either $T_{c1}$ or $T_{c2}$ in the two phase, $x_{c1} \leq x \leq x_{c2}$ region? The reason that superconductivity in pure $UBe_{13}$ needs to be compared to *both* superconducting states for $x_{c1} \leq x \leq x_{c2}$ may be seen in the phase diagram in Fig. 7. Specifically, if the fall in $T_c$ with increasing x in the one phase regime (one phase at least according to most specific heat results), $x \leq x_{c1}$, should join with a $T_c$ vs x dependence in $x_{c1} \leq x \leq x_{c2}$ that gives a smooth behavior of $T_c$ with composition, then the superconducting transition in $x_{c1} \leq x \leq x_{c2}$ that will be seen as the natural, smooth extension for the transition in $x \leq x_{c1}$ will be the *lower*, $T_{c2}$ transition. This agrees with the discussion of Lambert et al. [57] who, using ac susceptibility (which only sees the upper transition), found that the suppression of $T_c$ for $x<x_{c1}$ with pressure is at a rate of ~0.016 K/kbar. However, in the two phase, $x_{c1} \leq x \leq x_{c2}$, superconductivity with pressure behaves differently, and $T_{c1}$ is suppressed at a much faster rate of ~0.05 K/kbar. They therefore inferred that the upper transition in the two phase region $x_{c1} \leq x \leq x_{c2}$ in Fig. 7 is a different kind of superconductivity that for the composition region $x<x_{c1}$.

**$UBe_{13}$:**

In this summary section we consider what is the evidence, pro and con, for conventional superconductivity in $UBe_{13}$.

**Conventional:** Consistent with a conventional picture, as we have discussed above, UBe$_{13}$ has a low $\gamma_0$. Second, $C \propto T^3$, although cited by theory as evidence for line nodes, is in fact the temperature dependence long associated with the superconducting electronic specific heat, $C_{el}^{sc}$, in electron-phonon coupled superconductors (such as elements) before the advent of the BCS theory in 1957 caused experimentalists to try an exponential temperature fit to $C_{el}^{sc}$. Third, Overhauser and Appel, in 1985, made the observation [23] that the superconducting specific heat in UBe$_{13}$ can be scaled onto that of the known BCS, electron phonon coupled superconductor Pb. Thus, they argue that pure UBe$_{13}$ is a conventional, BCS superconductor.

**Unconventional:** In this discussion of the nature of the superconductivity in undoped UBe$_{13}$, what about the reports that there is an anomaly in the superconducting state (which would usually [16] be taken to imply unconventional superconductivity)? In addition to the discussion above of such an anomaly (either in the zero field specific heat, seen by ref. 33, or as a function of field in the specific heat, refs. 35-36) there are other measurements that bear on this question. Brison et al. [58], in their specific heat below $T_c$ in fields up to 7.9 T of a normal purity sample (single crystal or polycrystal not stated, $\Delta T_c$=0.09 K) report no anomaly in field similar to that seen by Ellman et al. [35], Figs. 12 and 13, and Kromer et al. [36] Instead, Brison et al. report a field-induced magnetic anomaly in their undoped UBe$_{13}$ indicated by an upturn in C/T below 150 mK which first appears at 1.9 T and a peak in C/T for 5.8 T at 105 mK. Shimizu et al. [59] report an anomaly in their dc magnetization measurements in the superconducting state as a function of field (e. g. at 2.6 T and 0.14 K) that they are unable to explain. Magnetic torque measurements by Schmiedeshoff et al. [60] also infer a field induced magnetic anomaly at low temperatures in UBe$_{13}$ above 3-5 T, as well as in the normal state (i. e. inconsistent with the work of Shimizu et al.) In agreement with the finding [35-36] of an anomaly below $T_c$ in field, Matsuno et al. [61] find an anomaly in single crystal UBe$_{13}$ in the surface impedance below $T_c$ at about 0.6$H_{c2}$.

To summarize, there is obviously a wealth of unusual behavior in pure UBe$_{13}$ in the superconducting state about which there appears to be little consensus. The result of Rauchschwalbe et al. [33], that there is an anomaly at about 0.5 $T_c$ in zero field in UBe$_{13}$ – unconfirmed despite there being many measurements of the specific heat of UBe$_{13}$ in the literature – may be due to issues of sample purity. The issue of whether there is a magnetic field induced anomaly below $T_c$ in pure UBe$_{13}$ (as seen in specific heat by Ellman et al. [35] and Kromer et al. [36] and in various other measurements [59-61]) has been of sufficiently inconsistent nature that it has attracted no conclusive theoretical explanation and remains open for understanding.

Considering now data of more conclusive nature, consider the power laws of various measurements in UBe$_{13}$. Consistent with the power law observed [46,50] from 0.2 up to 0.9 K in the electronic specific heat in the superconducting state, $C_{es} \propto T^3$, which implies an axial superconducting state with point nodes, penetration depth measurements [62] from $T_c$ 0.06 to 0.86 K found that $\Delta\lambda \propto T^2$. These data also implied an axial, point node p-wave symmetry. Further consistent with unconventional superconductivity in UBe$_{13}$, Han et al. [63] used Josephson tunneling between Ta and UBe$_{13}$ to infer a non-s-wave pairing symmetry in UBe$_{13}$.

Hiess et al. [64], using neutron scattering on a large single crystal of $UBe_{13}$, see a shifting of magnetic spectral weight when the sample becomes superconducting. This is the soi disant magnetic resonance [16] characteristic of unconventional superconductivity, considered to be strong evidence for a sign change in the superconducting energy gap at the Fermi surface. This would argue for s± or d-wave pairing symmetry, rather than triplet p-wave.

In conclusion, it would be surprising if $UBe_{13}$, with its enormous normal state specific heat γ of ~1000 mJ/molK$^2$ and its numerous unusual properties just enumerated above, would be a conventional superconductor. However, much of the data that are consistent with this conclusion are not as clear cut as, for example as will be discussed below, for $U_{1-x}Th_xBe_{13}$. Further measurement and theoretical insights would be useful.

**$U_{1-x}Th_xBe_{13}$:**

The existence of multiple superconducting phases in a material that is single phase in a structural sense is [16] prima facie evidence for unconventional superconductivity. There is no evidence that $U_{1-x}Th_xBe_{13}$ is not structurally single phase. Both $UBe_{13}$ and $ThBe_{13}$ exist in the cF112 (cubic, face centered) $NaZn_{13}$ structure; $UBe_{13}$ has a lattice parameter, $a_0$, of 10.256 Å and $ThBe_{13}$ has $a_0$=10.395 Å [5]. Doping Th into $UBe_{13}$ leads to a monotonic increase in $a_0$, linear with x (Vegard's law).

One reason the issue of a fourth phase line in the $T_c$ vs x phase diagram of $U_{1-x}Th_xBe_{13}$ (see the multiple phase diagrams under consideration in Figs. 7, 13-15) has been, and remains, of such interest is the work by Yip, Li, and Kumar [65]. They used thermodynamic constraints of general applicability to analyze polycritical points in a phase diagram. Their work was applied to $UPt_3$ but is valid for any polycritical point - like the one at $x_{c1}$ in $U_{1-x}Th_xBe_{13}$. Yip, Li and Kumar point out that if three phase boundary lines come together at a tricritical point, one of the phase boundaries must be first order - *or* there is a fourth, second order phase line intersecting the critical point. Since the data for $U_{1-x}Th_xBe_{13}$ are clear that none of the transitions are first order, there must indeed be a fourth, second order phase boundary. Although Zieve et al. [39], see Fig. 14, have found a fourth phase line under pressure, this remains an outstanding question in $U_{1-x}Th_xBe_{13}$, which presumably improved sample quality will help resolve.

In addition to the two phases in $U_{1-x}Th_xBe_{13}$ first observed by Ott et al. in the specific heat [11] (and first shown to be two superconducting phases by Rauchschwalbe et al.'s $dH_{c1}/dT$ data [32]), what other experimental findings bear on the nature of the two superconducting phases?

There is the pressure work of Lambert et al. [57], which shows that $T_{c1}$ in the composition region of two transitions is suppressed with pressure three times more rapidly than $T_c$ for $x<x_{c1}$. Lambert et al., who were measuring ac magnetic susceptibility and thus could not observe the variation of $T_{c2}$ with pressure, then inferred (see phase diagram, Fig. 7) that the lower transition for $x_{c1} \leq x \leq x_{c2}$ had the same nature as the transition for $x<x_{c1}$ and was therefore different in nature than the phase below $T_{c1}$. In long term annealing work (see Fig. 16), expected to remove non-magnetic defects, Kim et al. [12] found – in qualitative agreement with Lambert et al. – a distinct difference in the two transitions in $U_{0.97}Th_{0.03}Be_{13}$. They found that the lower anomaly in the specific heat, $\Delta C$, at $T_{c2}$ was much more enhanced with annealing than $\Delta C$ at $T_{c1}$, implying qualitatively that the lower transition was unconventional superconductivity and the upper was conventional. Consistent with this inference that the transition at $T_{c2}$ was unconventional and different from that at $T_{c1}$, Heffner et al. [29], and references therein, using µSR measurements, found that quasi-static magnetism, with a moment in the range of $10^{-3}$ to $10^{-2}$ $\mu_B$/U atom, appears below $T_{c2}$. Because of the limited temperature range available below $T_{c2}$ in $U_{1-x}Th_xBe_{13}$, there are not many power law determinations. Jin et al. [46], whose lowest temperature of measurement (0.09 K) was limited by the self-heat from the radioactive, depleted U, observed that - for all three of their measured concentrations of Th with $x>x_{c1}$ (see Fig. 9) – the low temperature specific heat can be fit from 0.09 to 0.32 K by a fully gapped BCS temperature dependence with an adjustable size of the energy gap, $\Delta$ - rather than $C\sim T^3$ as they found for pure $UBe_{13}$.

However, contrary to this long list of evidence that the two superconducting anomalies (at $T_{c1}$ and at $T_{c2}$) are different, Kim et al. [12] (in the low field (H≤1.25 T) limit) in the same work as the long term annealing result (Fig. 16) – using specific heat measurements - found that the rate of suppression in $U_{0.97}Th_{0.03}Be_{13}$ of $T_{c1}$ and $T_{c2}$ with applied field was the *same*. Jin et al. [43] measured the critical field of a similar high purity sample of $U_{0.97}Th_{0.03}Be_{13}$ to higher fields and found that the upper critical fields of both transitions was about the same, $H_{c2}(0)=4.8$ T. Since the upper transition at $T_{c1}$ starts at a higher temperature than the lower transition at $T_{c2}$, this work of Jin et al. could be interpreted as saying that the lower transition is more resistive to magnetic field than the upper one. Adding to the characterization of $U_{0.97}Th_{0.03}Be_{13}$ by measurement of the critical fields of the two transitions, Shimizu et al. [51] find in a single crystal that $T_{c2}(H)$ is isotropic whereas $T_{c1}$ is not.

### IV Theory and Conclusions:

Thus, in general (except for the low field data of Kim et al. [12]), various measurements [12,29,43,51, 56-57] find the lower superconducting phase below $T_{c2}$ in $U_{1-x}Th_xBe_{13}$ to be different than the upper one between $T_{c1}$ and $T_{c2}$. Various theories have been proposed to explain these fascinating and complex results. Kumar and Woelfle [66] proposed two different superconducting symmetries, d-wave at $T_{c1}$ and s-wave below $T_{c2}$, with a mixture of the two for $T_{c2} < T < T_{c1}$. In their theory, the muon result [29] of magnetism below $T_{c2}$, discovered later, is not addressed. Sigrist and Rice [41] also predict two superconducting symmetries being present, with non-unitary (unconventional) superconductivity below $T_{c2}$ (which would be consistent with the greater sensitivity [12] to annealing/removal of defects of the lower transition shown in Fig. 16), where such non-unitary pairing creates [16] a finite local spin polarization – consistent with the µSR result [29] of Heffner et al. There have long been discussions about whether the moment observed by µSR [29] is characteristic of a time reversal symmetry breaking unconventional superconducting transition at $T_{c2}$, or the occurrence of a spin density wave at around the same temperature due to some other effect. Unfortunately, despite the best efforts of Hiess et al. [67], neutron scattering has not been able to find definitive evidence pro or con to the question of the existence of some long range order at $T_{c2}$. In the case of Hiess et al., their detection limit was stated to be approximately 0.025 – 0.05 $\mu_B$, i. e. larger than the µSR [29] estimate of $10^{-3}$ to $10^{-2}$ $\mu_B$.

A non-unitary order parameter (proposed in the theory of Ohmi and Machida [68] and also as discussed by Sigrist and Rice) for $U_{1-x}Th_xBe_{13}$) below $T_{c2}$ would also be consistent with the large residual $\gamma_0$ in the low temperature superconducting state data of Jin et al. [46], where for a high purity, long term annealed sample of $U_{0.97}Th_{0.03}Be_{13}$ they found (see Table 2) $\gamma_0$=750 mJ/molK$^2$. This residual $\gamma_0$ is 30% of $\gamma_{normal}$ extrapolated [46] to T=0 from above $T_{c1}$ to match the superconducting and normal state entropies at $T_c$: $S_{sc}(T_c)=S_n(T_c)$

Quite recently, based on the results of Shimizu et al. [51,59], Machida [69] has proposed a new theoretical treatment of $U_{1-x}Th_xBe_{13}$, with a p-wave pairing symmetry (also [41,68] non-unitary, time reversal broken, unconventional order parameter) for $T<T_{c2}$, and a biaxial nematic phase in the 2D $E_u$ degenerate scenario for $T_{c2}≤T≤T_{c1}$. This model depends on the transition at $T_{c2}$ being time reversal symmetry breaking, consistent with the µSR [29] results. If in fact there is some long range SDW order, which as we have discussed is not ruled out by the neutron results [67], this would argue against the theory. A second possible reservation about the model is that it depends on the $H_{c2}$ vs T lines for $x_{c1}≤x≤x_{c2}$ not crossing. Such non-crossing is indeed consistent with known data (Kromer et al. [70]) but these data [70] for x=0.022 are incomplete. If one looks critically at the data of ref. 70 for the upper critical field for $T_{c2}$ for x=0.022, it looks rather clear that this upper critical field curve would cross that for $T_{c1}$ if only the data were taken lower than 0.2 K down to the work's otherwise lowest temperature of measurement of 0.1 K. A third possible point to consider is of course the result by Shimizu et al. [51] that $U_{0.97}Th_{0.03}Be_{13}$ is nodeless below $T_{c2}$ based on the lack of oscillations in C(H, T, $\Theta$) - a building block of the model [69] by Machida – has as discussed noise at the 3% level that could mask such oscillations. For example, in the C(H, T, $\Theta$) work by An, et al. [71] on CeCoIn$_5$, the amplitude of the oscillations seen was well below the 3% level.

To summarize the measured properties of UBe$_{13}$ and $U_{1-x}Th_xBe_{13}$, we consider these properties in light of the template in the recent review [16] of unconventional superconductivity. This template considers 16 separate properties that can bear on the nature of the pairing in a superconductor. As we have discussed, UBe$_{13}$ satisfies six of these criteria for unconventional behavior: large specific heat $\gamma$ ($\Rightarrow$ low [16] $T_{Fermi}$), C/T $\propto$ logT in the normal state ($\Rightarrow$ quantum critical behavior [17]), power laws (in the specific heat [46,50], C$\propto T^3$ $\Rightarrow$ point nodes; penetration depth [62] $\Delta\lambda \propto T^2$ $\Rightarrow$ point nodes); Josephson tunneling [63] $\Rightarrow$ non-s-wave, shift of neutron scattering magnetic spectral weight [64] ("magnetic resonance"), and $T_c$ is depressed by non-magnetic impurities (e.g. Al [47]) in a comparable fashion to magnetic impurities. Unsatisfied/under question criteria [16] that pertain to UBe$_{13}$ would include the existence of a second superconducting phase (but see Figs. 12-13, 15), time reversal symmetry breaking (see following discussion for Th-doped UBe$_{13}$), and C(H, $\Theta$) (done by Shimizu et al. [51] only for $U_{1-x}Th_xBe_{13}$. A number of the other 16 indications for unconventional superconductivity do not apply to either UBe$_{13}$ or $U_{1-x}Th_xBe_{13}$ (e. g. the existence of a pseudogap as commonly seen in the cuprate superconductors.)

$U_{1-x}Th_xBe_{13}$, for x between approximately 2 and 4 %, also shows clear indications of unconventional superconductivity: large γ, multiple superconducting phases, $T_{c2}$ and $\Delta C/T_{c2}$ are enhanced by annealing [12] (Fig. 16) – equivalent to the removal of non-magnetic impurities, the strong possibility of the breaking of time reversal invariance at $T_{c2}$ (see discussion above and the μSR [29] and neutron [67] data), and a large residual specific heat $\gamma_0$ in the superconducting state [46] (Table 2).

Although there can be no doubt that these two heavy Fermion superconductors exhibit unconventional superconductivity, there are outstanding questions about the pairing symmetries which would benefit from further work. Such work would include magneto-optic measurements [16] in $U_{1-x}Th_xBe_{13}$ of the Faraday or Kerr effects (one of the as-yet unsatisfied criteria for unconventional superconductivity from ref. [16]) to confirm whether time reversal symmetry is indeed broken at $T_{c2}$. Other useful measurements would include improving the noise level in the $C(H, T, \Theta)$ results in both $UBe_{13}$ and $U_{1-x}Th_xBe_{13}$.

As we have discussed, there are several competing theories to explain these unusual superconducting behaviors. Despite experimental and theoretical work that continues still, it seems difficult at this point to declare success in the attempt to understand the underlying mechanism and cause of these novel states. In an earlier theory review by Thalmeier and Zwicknagl [72], the extent of theoretical understanding of $U_{1-x}Th_xBe_{13}$ at the time was summarized "there is no developed microscopic theory for this complex behavior." Based on results since then, there are indeed new experimental data and new theoretical insights that offer hope that the precise nature of the unconventional superconducting lower transition and its pairing symmetry, as well as its cause, in $U_{1-x}Th_xBe_{13}$ will eventually be put into the solved column.

Acknowledgements: Helpful discussions with P. Kumar and assistance with the graphs by J. S. Kim and E.-W. Scheidt are gratefully acknowledged. Work performed under the auspices of the Bureau of Energy Sciences, U. S. Department of Energy, contract number DE-FG02-86ER45268.